\documentclass[prb,aps,amssymb,twocolumn,showpacs,floatfix]{revtex4}

\usepackage{graphics}
\usepackage{bm}

\usepackage{epsfig} 

\usepackage{dcolumn}

\usepackage{amsmath}

\begin{document}

\newcommand\dd{{\operatorname{d}}}
\newcommand\sgn{{\operatorname{sgn}}}
\def\Eq#1{{Eq.~(\ref{#1})}}
\def\Ref#1{(\ref{#1})}
\newcommand\e{{\mathrm e}}
\newcommand\cum[1]{  {\Bigl< \!\! \Bigl< {#1} \Bigr>\!\!\Bigr>}}
\newcommand\vf{v_{_\text{F}}}
\newcommand\pf{p_{_\text{F}}}
\newcommand\ef{{\varepsilon} _{\text{\sc f}}}
\newcommand\zf{z_{_\text{F}}}
\newcommand\zfi[1]{{z_{_\text{F}}}_{#1}}
\newcommand\av[1]{\left<{#1}\right>}

\title{Tunneling spectroscopy of  Luttinger-liquid structures
far from equilibrium}

\author{D. B. Gutman$^{1,2}$, Yuval Gefen$^3$, and A. D. Mirlin$^{4,1,2,5}$}
\affiliation{\mbox{$^1$Institut f\"ur Theorie der kondensierten Materie,
 Universit\"at Karlsruhe, 76128 Karlsruhe, Germany}\\
\mbox{$^2$DFG Center for Functional Nanostructures,
 Universit\"at Karlsruhe, 76128 Karlsruhe, Germany}\\
\mbox{$^3$Dept. of Condensed Matter Physics, Weizmann Institute of
  Science, Rehovot 76100, Israel}\\
\mbox{$^4$Institut f\"ur Nanotechnologie, Forschungszentrum Karlsruhe,
 76021 Karlsruhe, Germany}\\
\mbox{$^5$Petersburg Nuclear Physics Institute, 188300 St.~Petersburg, Russia}
}

\date{\today}

\begin{abstract}
We develop a theory of tunneling spectroscopy of interacting electrons
in a non-equilibrium quantum wire coupled to reservoirs. 
The problem is modelled as an out-of-equilibrium Luttinger liquid with
spatially dependent interaction. The interaction leads to the
renormalization of the tunneling density of states, 
as well as to the redistribution  of electrons over energies. 
Energy relaxation is controlled by 
plasmon scattering at the boundaries between regions with
different interaction strength,  
and affects the distribution function
of electrons in the wire as well as that of electrons
emitted from the interacting regions into non-interacting electrodes.
\end{abstract}



\maketitle

\section{Introduction}
\label{s1}

One-dimensional (1D) interacting fermionic systems show remarkable
physical properties and are promising elements for future
nanoelectronics.  The electron-electron interaction
manifests itself in a particularly dramatic way in 1D systems, inducing a
strongly correlated electronic state -- Luttinger liquid (LL)
\cite{stone,giamarchi,maslov-lectures,Delft}. 
A paradigmatic experimental realization of quantum
wires are carbon nanotubes \cite{zba-carbon-nanotubes}; for a recent
review see Ref.~\onlinecite{nanotubes}. Further
realizations encompass semiconductor, metallic and
polymer nanowires, as well as quantum Hall edges.

There is currently a growing interest in non-equilibrium phenomena on 
nanoscales. A  tunneling spectroscopy (TS) technique for non-equilibrium
nanostructures was developed in Ref.~\onlinecite{pothier97}. 
Employing  a superconducting tunneling electrode allows one to explore
not only the tunneling density of states (TDOS) but also the energy
distribution function. The energy relaxation found in this way
provides information 
about inelastic scattering in the system. In a very recent experiment
\cite{Birge}  this TS method was applied to a carbon nanotube 
under strongly non-equilibrium conditions.   

\begin{figure}[htbp]
\includegraphics[width=\columnwidth,angle=0]{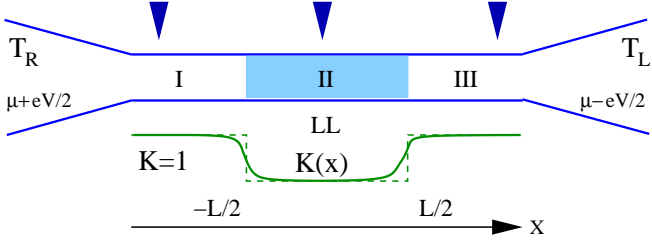}
\caption{Schematic view of a LL conductor with various positions of
tunnel probes.  The solid curve in the lower part of the figure shows a
spatially dependent LL interaction parameter $K(x)$. The dashed line
corresponds to the limit of a sharp variation of $K(x)$ at the boundaries.}
\label{fig1}
\end{figure}

In this paper, we develop a theory of TS of a LL  out of equilibrium.  
Specifically, we consider  a LL conductor connected, via non-interacting 
leads, to reservoirs with different 
electrochemical potentials, $\mu_L - \mu_R = eV$ and different
temperatures $T_L$, $T_R$ (where the indices $L$, $R$ stand for left-
and right-movers). It is assumed that the coupling to the leads is
adiabatic on the scale of the Fermi wave length, so that no
backscattering of electrons takes place. 
We model the leads as
non-interacting 1D wires, so that the electron-electron interaction
is turned on at the vicinity of the points $x=\pm L/2$, see
Fig.~\ref{fig1}. This model is quite generic to properly describe 
the problem at hand, independently of the actual geometry of the leads. 
Note also that the 1D setup with  strongly non-uniform 
interaction may be experimentally
realized by using external screening gates. 
 
It is known that  energy relaxation is absent in a uniform clean
LL. Within the golden-rule framework, the lack of energy relaxation 
for forward scattering processes results from 1D kinematic constraints
that do not allow to satisfy the
energy and momentum conservation laws simultaneously \cite{khodas}. 
On a more formal level,  the conservation of energies of individual particles 
in a spatially uniform LL  is protected by the integrability 
of the system, which implies an infinite number of conservation laws
\cite{mattis}. 
Inclusion of spatial dependence into the model   
violates these laws and leads to energy relaxation 
that takes place at the regions where the interaction varies in space
\cite{note-relaxation}.  

The fact that  inhomogeneous interaction induces  energy
relaxation of electrons has been pointed out for the first time 
in Ref.~\onlinecite{oreg95}
in the context of interacting quantum Hall
edges but a detailed analysis of this effect has been missing until now. 
On the other hand, one may  expect this to be a dominant effect on  
the electron distribution function in experiments 
done on modern high-quality quantum wires 
(such as ultraclean carbon nanotubes \cite{ultraclean-nanotubes}), 
under non-equilibrium conditions. There is thus a clear need in the theory
of TS in non-equilibrium LL.    

It is worth noting that we assume the absence of backscattering due to 
impurities in the wire. When present, such impurities strongly affect
the electronic properties of a LL wire: they induce diffusive dynamics at
sufficiently high temperature $T$ and  localization phenomena
proliferating with lowering $T$ (Ref.~\onlinecite{GMP}), as well as
inelastic processes \cite{Bagrets1,Bagrets2}. We also neglect the nonlinearity
of the electron dispersion whose influence on spectral and kinetic 
properties of 1D electrons was recently studied in
Refs.~\onlinecite{khodas}, \onlinecite{Matveev}.

\section{Formalism}
\label{s2}

Within the LL model, the electron field is decoupled in a sum of right-
and left-moving terms,  
$\psi(x,t)=\psi_R(x,t)e^{ip_Fx}+\psi_L(x,t)e^{-ip_Fx}$, where $p_F$ is
the Fermi momentum. The Hamiltonian of the system reads
\begin{eqnarray}&&
H=H_0+H_{\rm int}\,, \\&&
H_0= -iv \int dx \left(\psi_R^\dagger\partial_x\psi_R-
\psi^\dagger_L\partial_x\psi_L\right)\,, \\&&
H_{\rm int}= \frac{1}{2}\int dx g(x)(\psi_R^\dagger\psi_R
+ \psi_L^\dagger\psi_L)^2,
\end{eqnarray}
where  $v$ is the electron velocity and $g(x)$ is the 
spatially dependent electron-electron interaction constant. 

We will proceed by following the lines of the functional bosonization approach
\cite{func-bos} in the non-equilibrium (Keldysh) formulation
 \cite{Yur,GGM-2008}$^,$\cite{Bagrets1}. 
Performing the Hubbard-Stratonovich transformation, one decouples the
interaction term via a bosonic field $\phi$ and gets the action 
\begin{align}\label{TL}
S[\psi,\phi]=i\sum_{\eta=R,L}\psi^\dagger_\eta
(\partial_\eta-\phi)\psi_\eta-\frac{1}{2}\phi 
g^{-1}\phi\, ,
\end{align}
where $\partial_{R,L} = \partial_t\pm v\partial_x$ and the fields are
defined on the Keldysh time contour. 
The information about physical observables is contained in Keldysh
Green functions \cite{Kamenev} $G^>$ and $G^<$; see, in particular,  
Appendix \ref{s9} where we express tunneling current 
in terms of functions $G^\gtrless$ and  
discuss how its measurement  allows 
to determine $G^\gtrless$ experimentally. 
The Green functions $G^\gtrless$ can be presented in the form
\begin{eqnarray}
G^\gtrless_\eta(x,t;x',t') &=&
\int{\cal D}\phi\: Z[\phi] e^{-{i\over 2} \phi g^{-1}\phi} \nonumber\\
&\times& G^\gtrless_\eta[\phi](x,t;x',t'), 
\label{green}
\end{eqnarray}
where we introduced the Green function in a given field configuration, 
$G^\gtrless_\eta[\phi]$,  
and the sum of  vacuum loops, $Z[\phi]$.
 
In 1D geometry the coupling between the
fermionic and bosonic fields can be eliminated by a gauge
transformation 
$\nolinebreak{\psi_{\eta}(x,t)\to\psi_{\eta}(x,t)e^{i\theta_{\eta}(x,t)}}$,
if we require
\begin{eqnarray}&&
\label{diff_eq}
i\partial_{\eta}\theta_{\eta}=\phi\,.
\end{eqnarray}
As a result, $G_{\eta}^\gtrless[\phi]$ can be cast in the form
\begin{eqnarray}
G^{\gtrless}_\eta[\phi](x,t;x',t')&=& G^\gtrless_{\eta,0}(x-x';t-t')
e^{-i\eta eV(t-t')/2} \nonumber \\
&\times& e^{\Phi^\gtrless_\eta(x,t;x',t')} \,.
\end{eqnarray}
Here 
\begin{eqnarray}
\Phi^\gtrless_\eta(x,t;x',t')=
i\theta_{\pm,\eta}(x,t)-i\theta_{\mp,\eta}(x',t')\,, 
\label{Phi} 
\end{eqnarray}
$G^\gtrless_{\eta,0}$ is the Green function of free fermions,
\begin{eqnarray}
G^\gtrless_{\eta,0}(\xi)=\frac{T_\eta}{2v}
\frac{1}{\sinh\pi T_\eta(\eta\xi_\eta\pm i0)}\:, 
\end{eqnarray}
the coordinate $\xi_{R/L}=x/v \mp t$ labels the trajectory of a particle, 
and we use the convention that in formulas $\eta$ should be understood as
$\eta =\pm 1$ for right/left moving electrons.

It is convenient to perform a rotation in Keldysh space, thus
decomposing fields into classical and quantum components,
$\phi_1,\phi_2 = (\phi_+ \pm \phi_-)/\sqrt{2}$, where
the indices $+$ and $-$ refer to the fields  on two
branches of the Keldysh contour. Further, we 
introduce vector notations by combining $\phi_1$
and $\phi_2$ in a 2-vector $\bm{\phi}$. To proceed further, we resolve
Eq.~(\ref{diff_eq}) and express $\theta_\eta$ through $\phi$ as
\begin{equation}
\label{theta}
\bm{\theta}_\eta = {\cal G}_{\eta 0}\sigma_1 \bm{\phi}\,,
\end{equation}
 where ${\cal
  G}_{\eta 0}$ is the Green function of free bosons,
\begin{equation}
\label{green-function-bosons}
{\cal G}_{\eta 0} = \left(
\begin{array}{cc}
{\cal G}_{\eta 0}^K  & {\cal G}_{\eta 0}^r \\
{\cal G}_{\eta 0}^a & 0
\end{array}
\right)\,.
\end{equation}
Its retarded and advanced components are given by
\begin{equation}
\label{gra-bosons}
{\cal G}_{\eta 0}^{r,a} = {1 \over \omega-\eta v q \pm i0}
\end{equation}
The Keldysh component of ${\cal G}_{\eta 0}$ is given by ${\cal
  G}_{\eta 0}^K = ({\cal G}_{\eta 0}^r - {\cal G}_{\eta 0}^a)
B_\eta^{(0)}(\omega)$, where $B^{(0)}_\eta(\omega)$ is determined by
the temperature $T_\eta$ of the reservoir from which the electrons
moving in direction $\eta$ emerge,
\begin{eqnarray}&&
\label{e7}
B^{(0)}_\eta(\omega)=
\coth \omega/2T_\eta\,.
\end{eqnarray}
Using Eqs.~(\ref{Phi}) and (\ref{theta}) and performing a
transformation to the coordinate space, we express the exponent
$\Phi_\eta^\gtrless[\phi](x,t,x',t')$ 
through the bosonic field $\bm{\phi}(y)$, 
\begin{equation}
\Phi_\eta^\gtrless[\phi](x,t,x',t')= \int  {d\omega\over 2\pi} dy 
\bm{\phi}^T_{-\omega}
(y)\bm{J}^{\gtrless}_{\eta,\omega}(y;x,t,x',t').
\end{equation}
The components of $\bm{J}$ are found as
\begin{eqnarray}&&
\hspace{-0.5cm}J^{\gtrless}_{1,\eta,\omega}(y)
=\frac{e^{i\eta\frac{\omega}{v}y}}{\sqrt{2}v} 
\left(\theta[\!\eta(x\!-\!y\!)]e^{-i\omega\xi_\eta}\!
-\!\theta[\eta(\!x'\!-\!y\!)]e^{-i\omega\xi_\eta'}\right)\,,\nonumber 
\\&&
J^\gtrless_{2,\eta,\omega}(y)=-\frac{e^{i\eta
\frac{\omega}{v}y}}{\sqrt{2}v}
\left(e^{i\omega\xi_\eta}-e^{i\omega\xi_\eta'}\right)B_\eta^{(0)}(\omega)
\nonumber \\&& \mp
\frac{e^{i\eta\frac{\omega}{v}y}}{\sqrt{2}v}
\left(\theta[\eta(y-x)]e^{-i\omega\xi_\eta}
+\theta[\eta(y-x')]e^{-i\omega\xi_\eta'}\right),\label{j12}
\end{eqnarray}
where $\theta(x)$ is the Heviside $\theta$-function. 
The vacuum loop factor in Eq.~(\ref{green}) is given by
\begin{align}
Z[\phi] =\exp\left(-\frac{i}{2}\bm{\phi}^T \Pi \bm{\phi}\right)\,,
\end{align}
where $\Pi$ is the polarization operator,
\begin{eqnarray}&&   
\Pi\!=\!
\begin{pmatrix}
      0 & \Pi^a \\
      \Pi^r & \Pi^K
    \end{pmatrix}\:\!.
\label{polarization-operator-free} \nonumber
\end{eqnarray}
It can be decomposed into left and right moving parts, 
$\Pi=\Pi_R+\Pi_L$, 
with
\begin{eqnarray}&&
\label{J5}
\Pi^r_{R,L}=-\frac{1}{2\pi}\frac{q}{\omega_+\mp v_Fq}\,\, , \,\,
\Pi^a_{R,L}=-\frac{1}{2\pi}\frac{q}{\omega_-\mp v_Fq}\, ,  \nonumber \\&&
\Pi^K_\eta =(\Pi^r_\eta-\Pi^a_\eta)B^{(0)}_\eta(\omega)\,,
\end{eqnarray}
where $\omega_\pm =\omega\pm i0$. 
Performing the averaging over the auxiliary field $\phi$, we get 
\begin{equation}
\label{eq_G}
G^{\gtrless}_\eta(x,t;x',t')= G^\gtrless_{\eta,0}(\xi_\eta-\xi'_\eta)
e^{-i\eta eV(t-t')/2} e^{{\cal F}_\eta^\gtrless}\,, 
\end{equation}
where the effect of the interaction is represented by the ``Debye-Waller
factor'' $e^{{\cal F}_\eta^\gtrless}$ with
\begin{eqnarray}
{\cal F}_\eta^\gtrless (x,t;x',t') &=& -\frac{i}{2}\int {d\omega \over 2\pi}
dy_1dy_2 \nonumber \\
&\times & \bm{J}^{\gtrless,T}_{-\omega,\eta}(y_1){\cal V}_\omega(y_1,y_2)
\bm{J}^\gtrless_{\omega,\eta}(y_2)\,.
\label{f} 
\end{eqnarray}
Here 
\begin{eqnarray}
\label{int}
{\cal V}=(\Pi+g^{-1}\sigma_1)^{-1}
\end{eqnarray}
is the screened electron-electron interaction potential.  
Its retarded component is given by
\begin{eqnarray}
{\cal V}_\omega^{r}(y,y')=g(y)\bigg[\delta(y-y')+\frac{vg(y')}{\pi}
\partial_y\partial_{y'}
{\cal G}_\omega^r(y,y')\bigg]\,,
\end{eqnarray}
where the function  ${\cal G}_\omega^r$ is determined by the following
differential equation 
\begin{eqnarray} 
\label{plazmon}
(\omega^2+\partial_yu^2(y)\partial_y){\cal G}^r_\omega(y,y')=\delta(y-y')\,,
\end{eqnarray}
which describes the plasmon propagation  
in a medium with spatially dependent
sound velocity $u(x) = v(1+g(x)/\pi v)^{1/2}$. 
The Keldysh component of the interaction propagator is obtained as 
\begin{eqnarray}
{\cal V}_\omega^K(y_1,y_2)=-\frac{i\omega}{2\pi v^2}\sum_{\eta=\pm}
B_\eta(\omega) I^\eta_\omega(y_1)I^\eta_{-\omega}(y_2)\,,
\end{eqnarray}
where 
\begin{eqnarray}&&
I^\eta_\omega(y)=
\int dy'e^{i\eta\frac{\omega}{v}y'}{\cal V}_\omega^r(y,y')\,. 
\end{eqnarray}
At equilibrium, $B_R(\omega)=B_L(\omega)\equiv B(\omega)$, this 
reduces to
\begin{eqnarray}
{\cal V}_\omega^K=\bigg[{\cal V}_\omega^r-{\cal V}_\omega^a\bigg]B(\omega)\,,
\end{eqnarray}
in agreement with the fluctuation-dissipation theorem.

\section{Sharp boundaries}
\label{s3}

So far we made no restriction on the way the interaction changes in space.
Let us consider first the case when the interaction turns on and off
sharply on the scale set by the temperatures, $l_T \sim v/\max\{T_L,T_R\}$. 
This limit can be modelled via a stepwise interaction as  
represented by the dashed line in Fig.~\ref{fig1}.  Equation
(\ref{plazmon}) for ${\cal G}_\omega^r$ can be then straightforwardly
solved by using the fact that the velocity $u$ is constant in each of
three regions and employing the proper boundary conditions [continuity
of  ${\cal G}_\omega^r(y,y')$ and   
of  $u^2(y)\partial_y{\cal  G}_\omega^r(y,y')$] at $y=\pm L/2$. 

In the TS context, we are interested in the Green functions
$G^\gtrless$ with coinciding spatial arguments, $x=x'$. Assuming $x$
to be in the interacting part of the wire (and not too close to the
boundaries) and setting $t'=0$, we find    
\begin{eqnarray}&&
\label{eq_F}
\!{\cal F}_R^\gtrless=-\gamma\!\int_0^\infty\frac{d\omega}
{\omega}\bigg[\frac{(1-K)^2B_R^{(0)}(\omega)
+(1+K)^2B_L^{(0)}(\omega)}{2(1+K^2)}  
 \nonumber \\&&  
\times (1-\cos\omega t) \pm i\sin\omega t\bigg]\,,
\end{eqnarray}
where 
\begin{equation}
K = v/u \equiv (1+g/\pi v)^{-1/2}
\end{equation}
is the conventional dimensionless parameter characterizing 
the interaction strength in a LL and 
\begin{equation}
\gamma={(K-1)^2\over 2K}.
\end{equation}
The integral in Eq.~(\ref{eq_F}) and in analogous formulas
below is logarithmically divergent at large frequencies and require an
ultraviolet regularization. Specifically, these integrals are
understood as regularized by a factor $e^-{\omega/\Lambda}$, where
$\Lambda$ is an ultraviolet cutoff.
Deriving Eq.~(\ref{eq_F}), we have neglected terms of the form
$e^{i n\omega L/u}$ (with non-zero integer $n$)
that arise due to the Fabry-Perot-type interference of plasmon
modes reflected at the boundaries. Keeping these terms  
would lead to an additional oscillatory structure 
in energy \cite{Nazarov-97} with the scale $\pi u/L$.  
Since we  are interested in TS of long wires, we assume that this
scale is much less than $\max\{T_R,T_L\}$, so that oscillations are
suppressed. 

Substituting Eq.~(\ref{eq_F}) into Eq.~(\ref{eq_G}),  
we finally get the Green functions: 
\begin{equation}
\label{green_fun}
G^\gtrless_R(t)=(2\pi iv)^\gamma
\bigg[G^\gtrless_{R,0}(t)\bigg]^{1+\alpha}
\bigg[G^\gtrless_{L,0}(t)\bigg]^{\beta}e^{-i\eta eVt/2}\,,
\end{equation}
where 
\begin{eqnarray}
\alpha =\frac{(K-1)^4}{4K(1+K^2)}\ ,\ \ \ 
\beta=\frac{(K^2-1)^2}{4K(1+K^2)}\,.
\end{eqnarray}
The Green  functions (\ref{green_fun}) can be determined experimentally 
from  TS measurements \cite{Birge}, see Appendix~\ref{s9}. 
Their difference determines the 
TDOS  $\nu(\epsilon)$, 
\begin{eqnarray}
\label{App_nu}
G^>_{\eta}(\epsilon,x,x)-G^<_{\eta}(\epsilon,x,x)=-2\pi i\nu_\eta(\epsilon),
\end{eqnarray}
while each of them separately (or their sum) contains also 
information about the distribution function, as discussed below.
The results for the TDOS have been found in Ref.~\onlinecite{GGM-2008}. 

Next we consider the non-interacting parts of the wire, and discuss, e.g., the right moving electrons. 
In the region I (see Fig.~\ref{fig1}), 
$x, x' < -L/2$, we find from Eqs.~(\ref{f}), (\ref{j12}) that 
${\cal F}_R^\gtrless = 0$, so that the Green functions of the right movers
are not modified by interaction. Physically this is quite transparent: 
the right-moving electrons in this part of the system are just coming from
the reservoir and are not yet ``aware'' of the interaction with
the left-movers. 

The situation is distinctly different in the region III,
$x,x'>L/2$. Assuming $x=x'$,  we find
\begin{eqnarray}
\label{region3}
\!{\cal F}_R^{\gtrless} &=&
\int_0^\infty\frac{d\omega}{\omega} 
\frac{(1-K)^2}{1+K^2}
(1-\cos\omega t) \nonumber \\
&\times & [B^{(0)}_R(\omega)-B^{(0)}_L(\omega)]\,.
\end{eqnarray}
Substituting Eq.~(\ref{region3}) into Eq.~(\ref{eq_G}), one gets
\begin{eqnarray}
\label{eq2}
G^\gtrless_R(t)=\bigg[G_{R,0}^\gtrless(t)\bigg]^{\cal
  T}\bigg[G_{L,0}^\gtrless(t) 
\bigg]^{\cal R} e^{-i\eta eVt/2},
\end{eqnarray}
where 
\begin{eqnarray}
\label{tran_coeff}
{\cal T}=\frac{2K}{1+K^2}\,\,,\,\,\, {\cal R}=\frac{(1-K)^2}{1+K^2}\,.
\end{eqnarray}
Since ${\cal F}_R^\gtrless$ in Eq.~(\ref{region3}) is real, the TDOS
is not affected by the interaction, $\nu_R(\epsilon)=\nu_0 \equiv
1/2\pi v$, as
expected. The modification of the functions $G^\gtrless_R$ as compared
to that of incoming electrons, $G^\gtrless_{R,0}$, implies therefore
the change in the distribution function $n_R(\epsilon)$ of
right-movers. Indeed, for
non-interacting particles $G^<_R = 2\pi i \nu_0 n_R(\epsilon)$ and 
$G^>_R = - 2\pi i \nu_0 [1-n_R(\epsilon)]$.   
We thus see that the electrons ejected from the interacting part of
the wire into the lead are affected by the interaction: their
distribution function has changed. 

The left-moving electrons can be analyzed in the same way;
the corresponding results are obtained  by replacing  
$R\leftrightarrow L, V \leftrightarrow -V$ in  Eqs.~(\ref{green_fun}),
(\ref{eq2}). Clearly, the role of the regions I and III is
interchanged in this case. It is also worth mentioning that 
in the non-interacting parts of the wire the Green functions are both
Galilean and translationally invariant, depending on coordinates and
times via $\xi-\xi'$ only.

\section{Arbitrary boundaries}
\label{s4}

\begin{figure}[h]
\includegraphics[width=\columnwidth,angle=0]{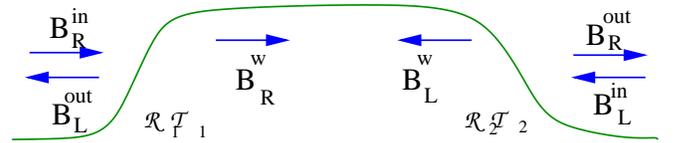}
\caption{
  Distribution functions of plasmons $B_\eta$ in different parts of
  the wire. The distributions of incoming plasmons are determined by
  respective leads, $B_\eta^{\rm in} = B_\eta^{(0)}$ } 
\label{fig2}
\end{figure}

We turn now to generalization of these results for the case of an
arbitrary shape of $g(x)$ in the contact region between the
interacting part of the wire and the non-interacting leads. The
contact regions are in general characterized by some reflection
coefficients $r_i(\omega)$ for the plasmon amplitude, yielding reflection
coefficients ${\cal R}_i=|r_i|^2$  for the plasmon intensity ($i=1,2$
for the left and 
right contact, respectively). The corresponding transmission
coefficients are ${\cal T}_i=1-{\cal R}_i$. 
It is instructive in this context to
compare our present approach with that developed in
Ref.~\onlinecite{GGM-2008}, where we analyzed the
tunneling density of states and focussed on the case of smooth
variation of $g(x)$ in the contact regions. As we are going to show,
the method of Ref.~\onlinecite{GGM-2008} can be generalized to the
case of arbitrary 
contacts (this was briefly discussed at the end of
Ref.~\onlinecite{GGM-2008}) and is also useful for the analysis of the electron
distribution function. Within that approach, the propagator of bosons
is calculated in momentum space (rather than in real space as in the
above calculation). The Keldysh component of the propagator is then
characterized by distribution function functions $B_\eta^{(0)}(\omega)$
and $B_\eta(\omega)$ associated with poles at $q=\eta \omega / v$ and $q=\eta
\omega / u$ and describing ``ghosts'' (free electron-hole pairs) and
plasmons, respectively \cite{note}.  
While the distribution function of ghosts is
simply determined by that of incoming electrons, the plasmons 
experience in general reflection at the boundaries. We have for the
left boundary (see Fig.~\ref{fig2})
\begin{equation}
\label{boundary-conditions}
B_R^{\rm w}={\cal T}_1B_R^{(0)}+{\cal R}_1B_L^{\rm w}\,,\qquad
B_L^{\rm out}={\cal R}_1B_R^{(0)}+{\cal T}_1B_L^{\rm w} \,,
\end{equation}
and similarly at the right boundary. Here we have introduced the notation
$B_\eta^{\rm w}$ for plasmon distributions in the interacting region
of the wire and $B_\eta^{\rm out}$ for out-going channels.
Solving these equations, we find the plasmon distribution functions of
right-movers in the interacting part of the wire, as well as in the
outgoing channel (in the right lead): 
\begin{eqnarray}
B_R^{\rm w}&=&\frac{{\cal T}_1}{1-{\cal R}_1{\cal
    R}_2}B_R^{(0)}+\frac{{\cal T}_2{\cal R}_1}{1-{\cal R}_1{\cal
    R}_2}B_L^{(0)}\,,  
\label{plas_kin}
\\
B_R^{\rm out} &=& \frac{{\cal T}_1{\cal T}_2}{1-{\cal R}_1{\cal R}_2}B_R^{(0)}+
\frac{{\cal T}_1+{\cal T}_2-2{\cal T}_1{\cal T}_2}{1-{\cal R}_1{\cal
    R}_2}B_L^{(0)}. 
\label{plas_kin_out}
\end{eqnarray}
The corresponding results for left movers are obtained by exchanging
the indices R$\leftrightarrow$L and 1$\leftrightarrow$2.

The method of Ref.~\onlinecite{GGM-2008} allows us to express 
the exponents ${\cal F}^{\gtrless}_\eta$  
in terms of these distribution functions. For the
interacting part of the wire, we get  
\begin{eqnarray}&&
\label{inf_theory_gen}
{\cal F}^{\gtrless}_R=-\int_0^\infty\frac{d\omega}{\omega}
\bigg[(B_R^{\rm w}-B_R^{(0)})
(1-\cos\omega t) \nonumber \\&&
+ \gamma\bigg((B_R^{\rm w}+B_L^{\rm w})(1-\cos\omega t)\pm 
i\sin\omega t\bigg)
\bigg]\,.
\end{eqnarray}
The result for the tunneling into the non-interacting region III of
Fig.~\ref{fig1}  can be obtained 
from Eq.~(\ref{inf_theory_gen}) by using the distribution
functions $B_\eta^{\rm out}$ corresponding to this region and
replacing the interaction constant $\gamma$ by zero,
 \begin{eqnarray}
\label{inf_theory_region3_gen}
&& {\cal F}^{\gtrless}_R\equiv {\cal F}_R=-\int_0^\infty\frac{d\omega}{\omega}
(B_R^{\rm out}-B_R^{(0)}) (1-\cos\omega t) \nonumber \\
&& =
\int_0^\infty\frac{d\omega}{\omega} 
{\cal R}
(1-\cos\omega t)[B_R^{(0)}(\omega)-B_L^{(0)}(\omega)]\,,
\end{eqnarray}
where ${\cal R}$ is the total reflection coefficient on a double-step 
structure, ${\cal R}= 1- {\cal T}_1{\cal T}_2/(1-{\cal R}_1{\cal R}_2)$.

For the case of sharp boundaries the reflection and transmission coefficients  
are given  by the Fresnel law,
${\cal R}_{1,2}=(1-K)^2/(1+K)^2$ and ${\cal T}_{1,2}=4K/(1+K)^2$, so that 
Eqs.~(\ref{inf_theory_gen}) and  (\ref{inf_theory_region3_gen}) reduce to the
earlier results  (\ref{eq_F}), (\ref{region3}). 
The total reflection and transmission
coefficients  ${\cal R}$ and ${\cal T}$ take in this case the values
(\ref{tran_coeff}) (which explains the notations introduced there). 
Clearly, the general formulas 
(\ref{inf_theory_gen}) and  (\ref{inf_theory_region3_gen}) can also be
obtained in the framework of a
 real-space calculation  that was
presented above for sharp boundaries. To do this,
one has to modify the boundary conditions for  the Green function in 
${\cal G}_\omega^r$ in Eq.~(\ref{plazmon}) by including the
appropriate reflection and transmission amplitudes $r_i(\omega)$ and
$t_i(\omega)$ at two boundaries and then proceeding in the same way as
in course of the derivation of Eqs.~(\ref{eq_F}) and
(\ref{region3}). The two methods (real space and k space)
are thus in full agreement
with each other. 

The formal results obtained thus far can be implemented to obtain physical 
observables. Consider first the
non-interacting part of the setup, region III of Fig.~\ref{fig1}.
The effect of the
interaction there amounts to modification of the distribution function
of outgoing particles (right-movers), which has (in time domain) the form
\begin{equation}
\label{eq_n}
n_R(t)=n_{R,0}(t)e^{{\cal F}_R(t)}\,,
\end{equation}
where ${\cal F}_R$ is given by Eq.~(\ref{inf_theory_region3_gen}).
This yields
\begin{eqnarray}
\label{n_R}
n_R(t) &=&
\frac{i}{2}e^{-ieVt/2}\left(\frac{T_R }{\sinh \pi T_R t+i0}\right)^{\cal
  T} \nonumber \\
&\times & \left(\frac{ T_L }{\sinh \pi T_L t+i0}\right)^{\cal R}.
\end{eqnarray}
The way in which the electron distribution function is modified depends 
on the kinetics of the plasmons inside the interacting region. 
For adiabatic switching of interaction, 
there is essentially no plasmon scattering. Therefore, the total 
reflection coefficient ${\cal R}$ and, consequently,  the 
exponent ${\cal F}_R$  in the region III 
vanish.  In this case the fermions retain their
distribution function: the right-movers going out into the right lead
have the same distribution as the right-movers injected into the
interacting region from the left lead. (The same applies to the
left-movers, of course.)  Let us now discuss  the opposite 
limit of strong reflection, ${\cal R} \rightarrow 1$. For a structure
with a sharp boundary, this is the case   provided the
interaction is strong, $K \rightarrow 0$. Alternatively, this limit
may be realized if the boundary regions are sufficiently extended and
characterized by random $K(x)$ such that plasmons with relevant
frequencies are localized. Regardless of  the cause, in the limit
${\cal R} \rightarrow 1$ the left- and right-moving electrons exchange their
distribution functions, except for keeping their total flux (i.e. the
chemical potential). 

Next, we consider the interacting part of the wire. Analyzing the
result (\ref{inf_theory_gen}), we see that two terms in square brackets
have distinctly different physical origin. The second term, which is
proportional to the local strength of the interaction $\gamma$ at the
measurement point is responsible for creation of the zero-bias anomaly (ZBA)
 as well as for its dephasing smearing, with the non-equilibrium dephasing rate
\cite{GGM-2008} 
\begin{eqnarray}
\label{tauphi}
\tau_\phi^{-1} &=& \pi\gamma \left[{(1-{\cal R}_1)(1+{\cal R}_2)\over
    1-{\cal R}_1{\cal R}_2} T_R \right. \nonumber \\ 
&+& \left.
{(1+{\cal R}_1)(1-{\cal R}_2)\over 1-{\cal R}_1{\cal R}_2} T_L\right]\,.
\end{eqnarray}
On the other hand, the first term in the integrand of
(\ref{inf_theory_gen}), which is governed by the difference between the
incoming and local distribution of plasmons, is fully analogous to the
expression for ${\cal F}^\gtrless$ in the non-interacting region,
Eq.~(\ref{inf_theory_region3_gen}), and describes the modification of the
distribution function inside the wire,
\begin{widetext}
\begin{eqnarray}
&& n_\eta(t) = n_{\eta,0}(t) \exp\left\{-\int_0^\infty {d\omega\over
    \omega} [B_\eta^{\rm w}(\omega) - B_\eta^{(0)}(\omega)] (1-\cos\omega
  t)\right\} \nonumber \\
&& =\frac{i}{2\pi}\frac{1}{t+i0} \exp\left\{-\int_0^\infty {d\omega\over
    \omega} [B_\eta^{\rm w}(\omega) - 1] (1-\cos\omega t)\right\}\,.
\label{distr_func}
\end{eqnarray}
\end{widetext} 
As is clear from Eq.~(\ref{distr_func}), the ``ghost'' term with
$B_\eta^{(0)}$ essentially serves to cancel the bare distribution
function  $n_{\eta,0}$, so that the distribution function $n(t)$
is determined only by the plasmonic distribution $B_\eta^{\rm w}(\omega)$ in
the wire. This is in fact a manifestation of a general relation
between the functional and full bosonization approaches, as will be
discussed in detail elsewhere \cite{tobe}.

\begin{figure}[h]
\includegraphics[width=\columnwidth,angle=0]{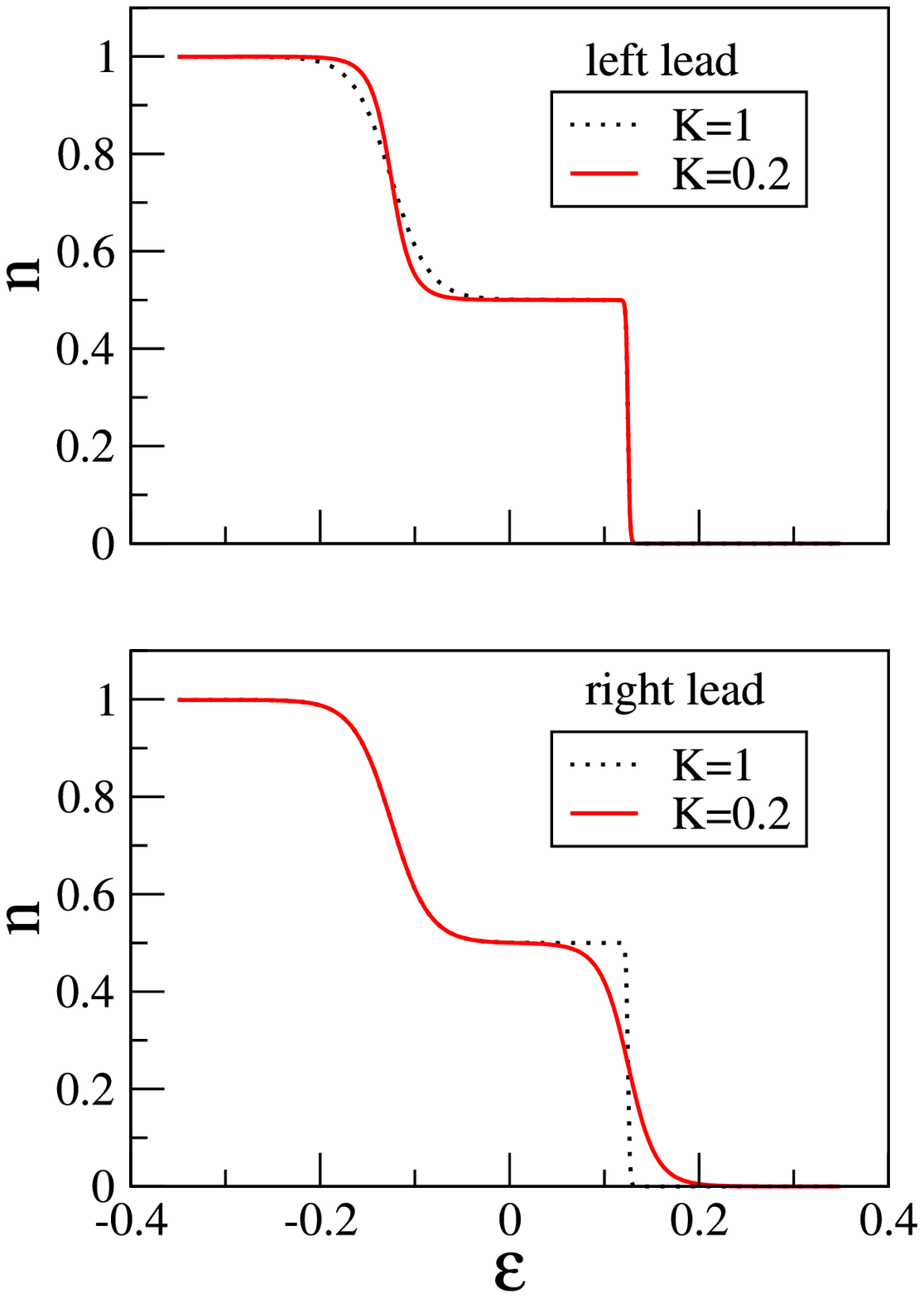}
\caption{Total distribution functions of electrons in the left 
and right leads for the LL interaction parameters $K=1$ (no
interaction) and $K=0.2$ (with sharp boundaries).
Temperatures of the leads are $T_L=0.2$ and
$T_R=0.001$; the bias voltage is $eV=0.25$.}
\label{fig3}
\end{figure}

\begin{figure}[h]
\includegraphics[width=\columnwidth,angle=0]{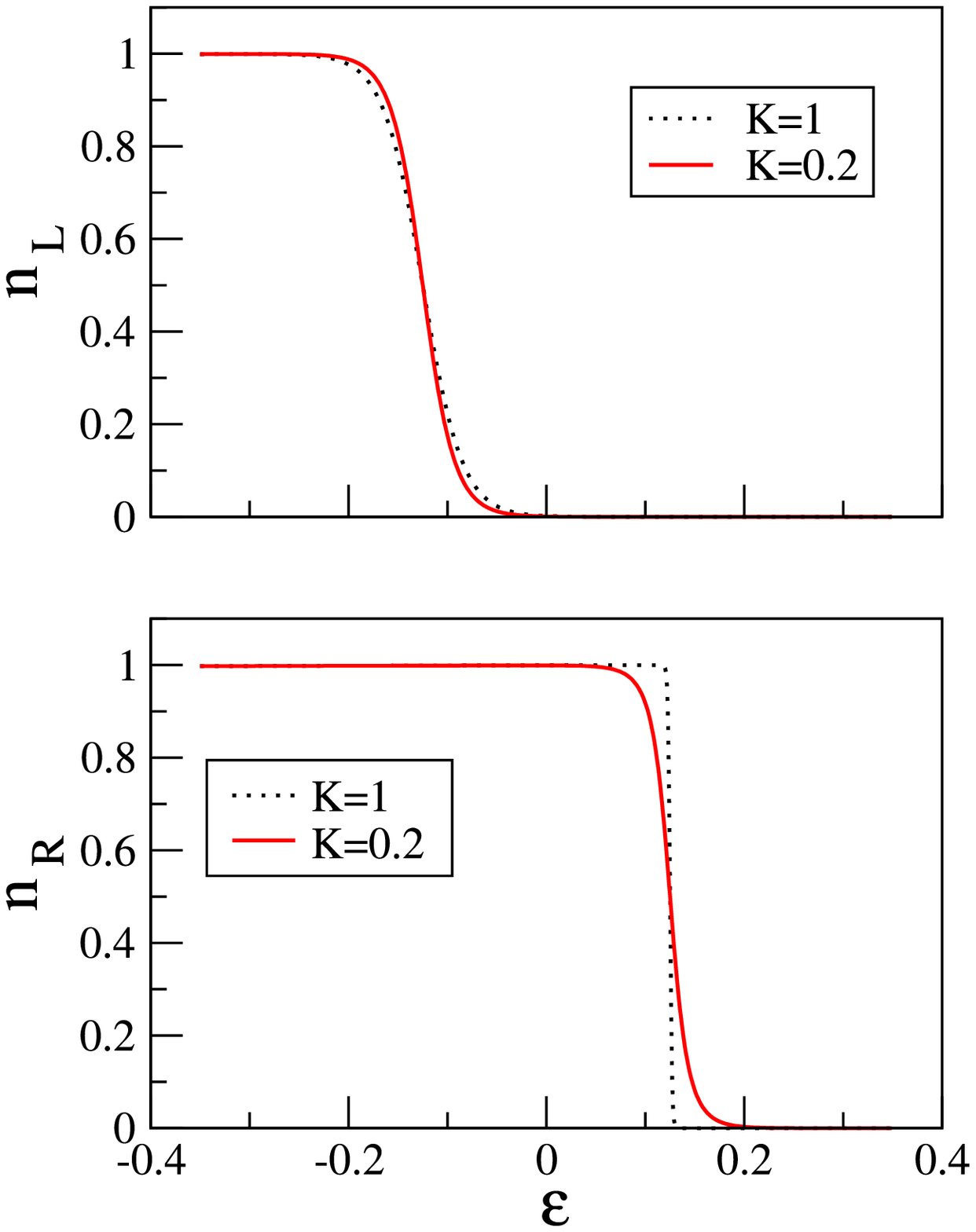}
\caption{Distribution functions of left- and right-movers,
  Eq.~(\ref{distr_func}), in the interacting part of the wire. All
  parameters are the same as in Fig.~\ref{fig3}.} 
\label{fig4}
\end{figure}

Fourier transformation of our results into the energy representation
can be done numerically (for analytic  calculation at  equilibrium see  
Appendix \ref{s8});
representative results are shown in
Figs.~\ref{fig3}, \ref{fig4}.  
In Fig.~\ref{fig3} we present distribution functions for 
non-interacting parts  of the wire.  
Temperatures are set to $T_L=0.2$ and $T_R=0.001$ (in arbitrary units), 
the applied voltage is $eV=0.25$, and a sharp variation of the interaction at the boundaries  (as in Sec.~\ref{s3}) is assumed.
 The distribution function of free
fermions  ($K=1$),  plotted by a dashed line, is the same
on both ends of the wire.
For interacting electrons (we choose the interaction parameter to be 
$K=0.2$, which is in the range of characteristic values reported for
carbon nanotubes, see, e.g., Ref.~\onlinecite{zba-carbon-nanotubes})
the distribution functions  in two leads are different. 
In particular, the distribution function in the left lead (region I in
Fig.~\ref{fig1}) has a sharp edge at the energy $\epsilon=\mu+eV/2$, which
corresponds to cold right-moving electrons.   
In the right lead (region III),  this edge is  broadened due to
interaction with hot  left-moving electrons. 
The situation is opposite for left-moving particles.
The distribution in the right lead has a broad edge at $\epsilon=\mu-eV/2$ 
that corresponds  to hot left-moving electrons.  
Due to interaction inside the wire this edge in the region I sharpens.

In Fig.~\ref{fig4}  we present the results for the distribution
functions of left- and right-moving quasiparticles in the central
(interacting) part of the wire, Eq.~(\ref{distr_func}). 
For $K=0.2$ the plasmon reflection at the boundaries is strong. 
In a symmetric structure this leads to almost equal distribution
functions of both types of carriers inside the wire.

\begin{figure}[h]
\includegraphics[width=\columnwidth,angle=0]{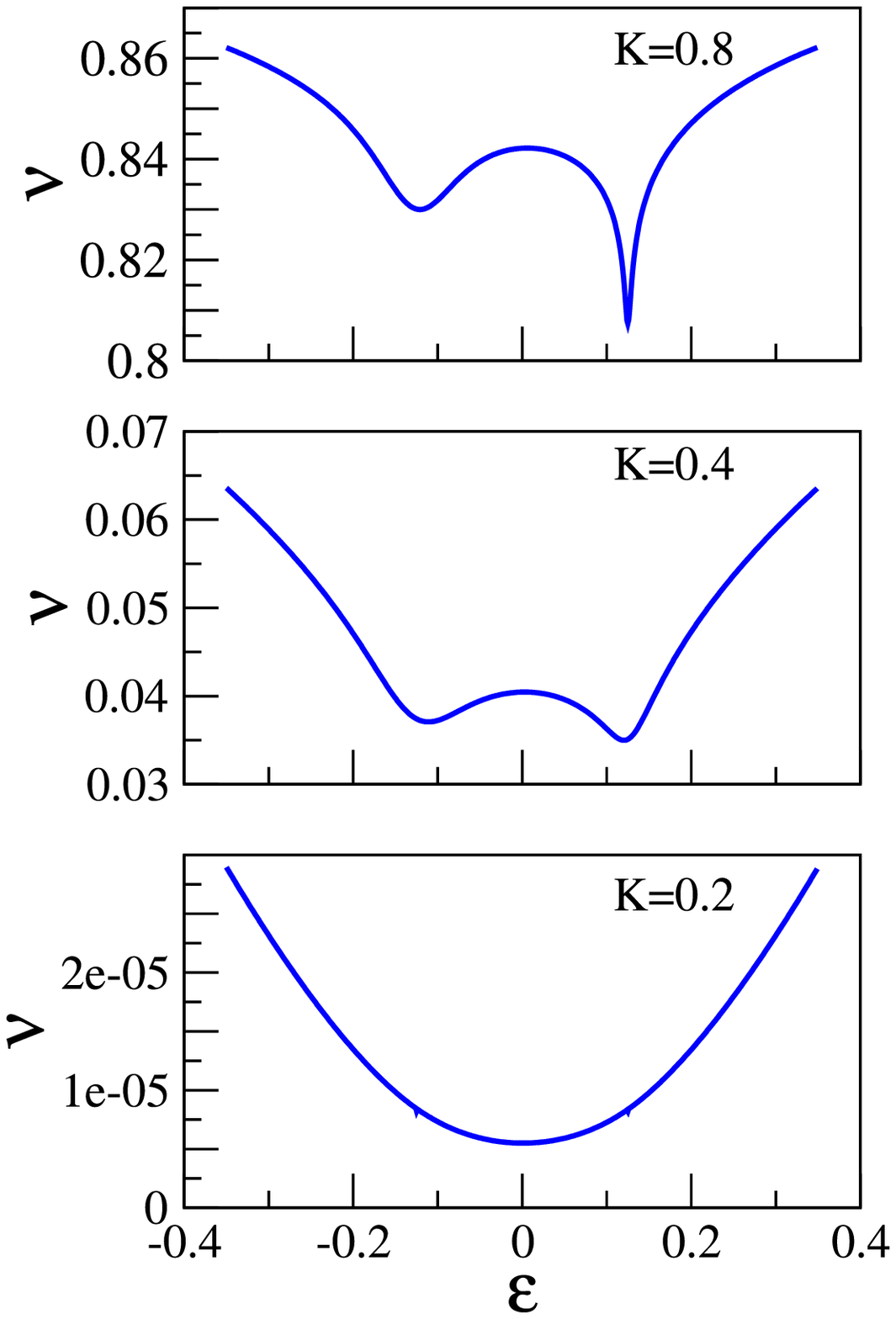}
\caption{TDOS $\nu(\epsilon)$ (normalized to its non-interacting value
  $2\nu_0$) in the interacting region for $K=0.8$, 0.4, and 0.2. The
temperatures of leads and the voltage are the same as in Fig.~\ref{fig3}.}
\label{fig5}
\end{figure}

In the upper panel of 
Fig.~\ref{fig5} we show the results for TDOS for $K=0.8$. 
The minima of TDOS are reached at energies $\epsilon=\mu\pm eV/2$. 
The broadening of the ZBA dips has two origins: smearing of the 
distribution function and dephasing. While the dephasing broadening
[cf. second term in Eq.~(\ref{inf_theory_gen})] is the same for both
chiral branches, the distribution functions [cf. first term in
Eq.~(\ref{inf_theory_gen})] are in general different.   
A deeper minimum at $\epsilon=\mu+eV/2$ reflects the fact that 
right-moving electrons in the wire have a much narrower distribution
function. This is because at $K=0.8$ the energy relaxation at the 
boundaries is quite weak, so that the distribution functions of cold
right-movers and hot left-movers are only slightly modified.
The situation is different for $K=0.2$, when distribution functions
$n_R$ and $n_L$ are nearly identical (up to a shift by $eV$), 
see Fig.~\ref{fig4}. 
As a result, the structure of the TDOS also becomes
symmetric. In fact, for the chosen value of the voltage, two broad ZBA
dips merge together.

\section{Thermal conductivity and electronic distribution function}
\label{s5}

We discuss now a relation between our results for the electron distribution
function and previous findings on the electric and thermal conductance of a
LL wire. In  the absence of backscattering the
number of left and right moving particles is separately preserved.
As a result, the  electric current is linear in the voltage $V$,
\begin{equation}
I=ev(N_R-N_L)=\frac{e^2}{\hbar} V
\end{equation}
with unrenormalized Landauer conductance $G=e^2/h$, Ref.~\onlinecite{Maslov}.
In our formalism,  this relation immediately follows from Eq.~(\ref{eq_n})
and the condition ${\cal F}_\eta(t\rightarrow 0) \rightarrow 0$. This
ensures that 
the modification of the distribution function of right (or left) movers by a
spatially varying interaction does not affect the integral of the distribution
function over energy, i.e. the total number of carriers of each type.

We turn now to the thermal conductance. The energy current is easily
found from the Green functions of electrons in non-interacting parts
of the wire,
\begin{equation}
\label{thermal_0}
I_E=\left. v\partial_t\left[G^<_R(t,t')-G^<_L(t,t')\right]\right|_{t=t'},
\end{equation} 
which
can be rewritten in terms of the electron distribution functions,
\begin{eqnarray}
\label{thermal_1}
I_E=\int_{-\infty}^\infty\frac{d\epsilon}{2\pi} \epsilon 
[n_L(\epsilon)-n_R(\epsilon)]\,.
\end{eqnarray}
Substituting the result (\ref{eq_n}), (\ref{inf_theory_region3_gen})
for the distribution functions, we get the expression of the thermal
current in terms of distribution functions of incoming electron-hole
pairs,
\begin{equation}
I_E 
= \frac{1}{4\pi}\int_0^\infty d\omega\omega{\cal T}(\omega)
[B_L^{(0)}(\omega)-B_R^{(0)}(\omega)]\,.
\label{thermal_2}
\end{equation}
According to (\ref{thermal_2}), the thermal conductance is affected by the
interaction [through the reflection coefficient ${\cal T}(\omega)$], as was
first found in  Ref.~\onlinecite{Fazio}. 
Note that due to the particle-hole symmetry of LL model,
the applied voltage  drops out of Eq.~(\ref{thermal_2}).
For the case of sufficiently sharp
boundaries, when  ${\cal T}(\omega)$ can be considered as $\omega$-independent
for relevant frequencies, Eq.~(\ref{thermal_1}) reduces to  
\begin{eqnarray}
I_E=\frac{\pi}{12}{\cal T}(T_R^2-T_L^2)\,.
\label{thermal_3}
\end{eqnarray}
Deviation of the transmission coefficient ${\cal T}(\omega)$ 
from unity leads to  the violation of the Wiedemann-Franz law
\cite{Fazio}. As is seen from our 
analysis, this deviation is a manifestation of a microscopic phenomenon:
energy relaxation of electrons due to non-uniform interaction.  

The heat current (\ref{thermal_2}) can be equivalently represented in
terms of plasmonic distributions in the wire 
\begin{equation}
I_E = \frac{1}{4\pi}\int_0^\infty d\omega\omega [B_L^{\rm
  w}(\omega)-B_R^{\rm w}(\omega)]\,.
\label{thermal_4}
\end{equation}
This implies that the presentation of the heat current in the form
(\ref{thermal_1}) is also valid in the interacting part of the wire,
with the electronic distribution functions $n_\eta(\epsilon)$
given by (\ref{distr_func}). Thus, also in the interacting part of the wire,
the energy current can be understood as carried  by properly defined
quasi-particle excitation. This is a remarkable result, which demonstrates that
the concept of fermionic quasiparticles remains meaningful in a 
strongly interacting 1D system (LL) despite its non-Fermi-liquid features. 

\section{Summary and outlook}
\label{s6}

To summarize, we have developed a theory of tunneling spectroscopy 
of  LL conductor connected to reservoirs away from equilibrium. 
In the specific setup considered here, each branch originates from a
source which is at equilibrium. However, the right and the left
sources have different temperatures and different chemical
potentials. 
We have modeled the system as a LL with spatially non-uniform
interaction, and calculated the single-electron Green functions
$G^\gtrless$ that carry information about the TDOS and the 
fermionic distribution functions in different 
parts of the wire. The interaction affects the tunneling
characteristics in three distinct ways. First, it induces a
power-law ZBA in the TDOS $\nu(\epsilon)$ 
(with two dips split by the voltage) in the interacting part of the
wire.  Second, it leads to broadening of ZBA singularities due to
dephasing, with the dephasing rate governed by the interaction
strength and the plasmon distribution inside the wire. Both the ZBA and
the dephasing effects are encoded in the second term of
Eq.~(\ref{inf_theory_gen}). 

The third effect of the interaction---which is specifically at 
the focus of the present work---is the 
inelastic scattering of electrons, leading to
their redistribution over energies. This effect takes place in those
regions where the interaction strength varies in space 
(near the wire boundaries in our model), inducing
backscattering of plasmons (but not of electrons). 
This leads to relaxation of the electron
distribution functions: left and right moving fermions ``partly exchange''
their distributions, see Eqs.~(\ref{n_R}), (\ref{distr_func}) and
Figs.~\ref{fig3}, \ref{fig4}, \ref{fig5}. 
For slowly varying interaction, when the plasmons with relevant
frequencies go through essentially without reflection, the energy
relaxation of electrons is negligible. In the opposite limit, when the
plasmons are almost entirely reflected (due to strong and sharply
switched interaction or, else, due to disordered boundary regions
inducing the plasmon localization), the left- and right-movers essentially
exchange their distribution functions (but not their total
density). We have also discussed a connection between these results and
earlier findings on the thermal conductivity of LL structures. 

Our results are important for the analysis of TS experiments on
strongly correlated 1D structures (in particular, carbon nanotubes
\cite{Birge})  
out of equilibrium. In this connection, let us emphasize the following
important point. What can actually be measured in experiment are Green
functions, 
 $G^>$ and $G^<$. The TDOS $\nu(\epsilon)$ in the interacting part of
 the wire, as well as the distribution function $n(\epsilon)$ in the
 non-interacting regions are related to $G^>$ and $G^<$ in a simple
 way.  
On the other hand, in order to extract  
the distributions $n_R(\epsilon)$ and $n_L(\epsilon)$  from 
$G^\gtrless$  in the {\it interacting} part of the wire, a non-trivial
deconvolution procedure is necessary. The broadening  of (split)
Fermi-edge structures 
in $G^\gtrless$ in the interacting part of the wire is governed by both
the distribution function and the  dephasing. 
The dephasing contributes to the smearing of Fermi-edge singularities
also in higher-dimensional (diffusive) systems \cite{GGM-dif}, and 
should be taken into account for the accurate interpretation of 
corresponding experiments\cite{pothier97,Anthore}.
In the 1D case the role of dephasing becomes particularly dramatic (if
the interaction is sufficiently strong). This is very well illustrated
by 
Fig.~\ref{fig5}: two Fermi-edge singularities almost  (middle panel)
or even completely 
(lower panel) merge, despite the fact that the Fermi edges in the
distribution functions remain well separated (Fig.~\ref{fig4}). 

A comment of a more general nature is in order here. 
Our results illustrate the fact that there is no unique answer to the
question: ``How much is a LL different from a Fermi liquid?'' On one
hand, the strong, power-law ZBA in TDOS of a LL clearly distinguishes
it from the Fermi liquid. In more formal terms, the single-particle residue
$Z$, which is finite in the Fermi liquid, vanishes in a power-law
fashion at the Fermi level of the LL. Also the dephasing rate
determining the broadening of ZBA, Eq.~(\ref{tauphi}), is linear in
temperature, contrary to the Fermi-liquid $T^2$ behavior. One could
think that it makes little sense to speak about fermionic excitations
in this situation, but this is not the case. First, the power-law
vanishing of TDOS has little importance (like the value of $Z$ in the
Fermi liquid) for kinetic properties of the system. Second, the
dephasing rate (\ref{tauphi}) is governed by processes with zero
energy transfer and do not lead to any energy relaxation. As a result,
the distribution function of fermionic excitations,
$n_\eta(\epsilon)$, is a fully meaningful concept even in the case of
a strong interaction. It stays preserved as long as the interaction is
spatially constant (or varies adiabatically slow with
$x$). Furthermore, both the charge and the energy current in the
interacting part of the wire can be
understood as carried by these fermionic quasiparticles.     
From this point of view, the LL is a {\it perfect} Fermi liquid. 

We conclude the paper by reviewing some future research prospects; the work
in those directions is currently underway. First, one may consider a
more general non-equilibrium situation where the distribution functions
``injected'' into the interacting part of the wire are of
non-equilibrium (e.g., double-step) form by themselves
\cite{GGM-2008,Jacobs}, see setups b, c in Fig.1 of Ref.~\onlinecite{GGM-2008};
the first of these setups is close to the
experimental situation of Ref.~\onlinecite{Birge}.
This requires a generalization of the bosonization technique 
that will be presented elsewhere \cite{tobe}. 
Second, it is interesting to study correlations
between outgoing left- and right-movers. In a general situation, one
finds that their density matrices are not decoupled, i.e. they are
entangled, which manifests itself, in particular, in
current cross-correlations. Third, one may study the effect of a
random variation of the interaction strength $K(x)$ in the wire. If
the wire is sufficiently long,  plasmons with not too low frequencies
get localized. Using our general results, one concludes that in the
left (right) half of the wire both distributions $n_R$, $n_L$ are determined
by that of the left (respectively, right) reservoir, with a transition
region which extends over the localization length of the middle section.
To refine this picture, one has to include into consideration
also plasmons with low frequencies (that remain delocalized). Also,
including the spectral curvature will induce plasmon decay
processes. (In the context of thermal conductivity, this physics was
discussed in Ref.~\onlinecite{Fazio}.) Finally, our results can be
generalized to 
the case of chiral LL, where both branches move in the same direction,
which is the situation characteristic for quantum-Hall edge-state
devices \cite{edge-states}.

\section{Acknowledgments}
\label{s7}

We thank  
D. Bagrets, N. Birge,  A. Finkelstein, I. Gornyi, D. Maslov,
Y. Nazarov, D. Polyakov, and R. Thomale  for useful discussions. 
We are particularly grateful to the late
Yehoshua Levinson for numerous illuminating discussions on the physics
of non-equilibrium systems. This work was supported by
 US-Israel BSF, ISF of the Israel
Academy of Sciences, the Minerva Foundation, 
and DFG SPP 1285 (YG),
EC Transnational Access Program at the
WIS Braun Submicron Center (ADM), German-Israeli Foundation under Grant 965, 
and Einstein Minerva Center.

\appendix
\section{measurement of Green functions $G^\gtrless$}
\label{s9}
The tunneling current between a probe and a quantum wire 
can be expressed  in terms of the functions  $G^\gtrless$ as
\begin{eqnarray}
\label{App_A_eq1}
I(U) &=& \int dy dy' |T_{y,y'}|^2\int\frac{d\epsilon}{\pi} \nonumber \\
&\times &
\bigg[
G^<_{{\rm tp}}(\epsilon-eU,y,y')G^>_{{\rm w}}(\epsilon,y',y)\nonumber \\
&-& G^>_{{\rm tp}}(\epsilon-eU,y,y')G^<_{{\rm w}}(\epsilon,y',y)\bigg]\,,
\end{eqnarray}
where the subscripts ``tp'' and ``w'' refer to the tunnel probe 
and the wire respectively,  $U$ is a voltage between 
the tunneling probe  and the wire, 
and $T(y,y')$ is a tunneling matrix element in the coordinate representation.
If electron tunneling is local in space, we have  
$T(y,y')=T\delta(y-y')\delta(y-x)$, 
where $x$ is  a position of tunneling probe. 
Since the tunneling probe is at equilibrium, one can use a standard 
relation between the Green functions and distribution function 
$n_{\rm tp}(\epsilon)$  of electrons in the probe,

\begin{eqnarray}&&
\label{App_A_eq3}
G^<_{\rm tp }(\epsilon,x,x)=2\pi i\nu_{\rm tp}(\epsilon)n_{\rm
  tp}(\epsilon), \nonumber \\&& 
G^>_{\rm tp}(\epsilon,x,x)=-2\pi i\nu_{\rm tp}(\epsilon)
[1-n_{\rm tp}(\epsilon)].
\end{eqnarray}

Differentiating the tunneling current with respect to voltage and substituting 
Eq.~(\ref{App_A_eq3}) into Eq.~(\ref{App_A_eq1}), one finds
\begin{widetext}
\begin{eqnarray}&&
\label{App_eq_2}
\frac{\partial I}{\partial U}=-2i |T|^2\int d\epsilon\bigg\{
\frac{\partial \nu_{{\rm tp}}(\epsilon-eU)}{\partial \epsilon}
\bigg[
n_{{\rm tp}}(\epsilon-eU) G^>_{{\rm w}}(\epsilon,x,x)+
(1-n_{{\rm tp}}(\epsilon-eU))G^<_{{\rm w}}(\epsilon,x,x)
\bigg] 
\nonumber \\&&
-2\pi i\nu_{{\rm tp}}(\epsilon-eU)\nu_{{\rm
    w}}(\epsilon)\frac{\partial n_{{\rm tp}}(\epsilon-eU)}{\partial
  \epsilon}\bigg\} \,. 
\end{eqnarray}
\end{widetext}
For a LL wire the Green functions $G_{\rm w}$ and the TDOS $\nu_{\rm
  w}$ represent a sum of contributions  of both chiral branches. 
If the density of states in  the tunneling probe ($\nu_{{\rm tb}}$)  
is a constant  (as in a normal metal), the first term in 
Eq.~(\ref{App_eq_2})
drops out. In this case the result is proportional to the TDOS  in the wire.
Assuming  that the  tunneling probe is at zero temperature,  one then finds  
\begin{equation}
\frac{\partial I}{\partial U}=4\pi|T|^2\nu_{{\rm tp}}\nu_{{\rm w}}(eU).
\end{equation}
On the other hand, if the density of states in the tunneling probe is
strongly  
energy dependent (as for superconducting electrodes),  
the first term in Eq.~(\ref{App_eq_2}) survives. 
Unlike TDOS (which is determined by the difference 
$G_{\rm w}^>-G_{\rm w}^<$), this term contains also the information about 
$G_{\rm w}^>+G_{\rm w}^<$.
Therefore,  measurement  of the  tunneling current 
with two different types  of  tunneling probes (normal  and superconducting) 
allows one  to find functions $G^>_{\rm w}$ and $G^<_{\rm w}$ separately.
The idea to use superconducting electrodes for the tunneling
spectroscopy was introduced in Ref.~\onlinecite{pothier97} and more
recently  employed in  
Ref.~\onlinecite{Birge}.

\section{Green functions $G^\gtrless$ at thermal equilibrium}
\label{s8}
At thermal equilibrium  the Green functions in 
the energy domain can be calculated explicitly.
Using Eq.~(\ref{inf_theory_gen}) and 
$B_R^{\rm w}=B_L^{\rm w}=B_R^0=B_L^0=\coth\frac{\omega}{2T}$, we find 
the exponent ${\cal F}_\eta(t)$ for the Green functions in interacting 
part of the wire,
\begin{eqnarray}
{\cal F}(t)=\gamma\log\frac{\pi T}{i\Lambda\sinh\pi T(t-i/\Lambda)}\,, 
\end{eqnarray}
where we drop the chirality index $\eta$, as it is immaterial for $x=x'$ 
in equilibrium.
Using  Eq.~(\ref{eq_G}) and performing a Fourier transform from the 
time into the energy domain,  one finds
\begin{equation}
G^>(\epsilon)=-\frac{(\pi T)^{1+\gamma}}{2\pi v (i\Lambda)^\gamma}
\int_{-\infty}^{\infty} dt
e^{i\epsilon t}\frac{1}{\sinh^{1+\gamma}\pi T(t-i/\Lambda)}\,.
\end{equation}
After calculating  an  auxiliary integral 
\begin{eqnarray}&&
\int_{-\infty}^{\infty}dt
\frac{e^{izt}}{\sinh^{1+\gamma}(t-i0)}=\nonumber \\&&
\times\frac{i^{1+\gamma}2^\gamma}{\Gamma(1+\gamma)}e^{\pi z/2}
\Bigl|\Gamma[(1+\gamma+iz)/2]\Bigr|^2 \,,\nonumber
\end{eqnarray}
one obtains
\begin{equation}
\label{App3}
G^>(\epsilon)=-\frac{i}{2\pi v} 
\frac{2^\gamma}{\Gamma(1+\gamma)}
\left(\frac{\pi T}{\Lambda}\right)^\gamma 
e^{\pi z\over 2}\Bigl|\Gamma[(1+\gamma+iz)/2]\Bigr|^2 ,
\end{equation}
where $z=\epsilon/\pi T$. 
Similarly, one finds  the function $G^<$,
\begin{equation}
\label{App4}
G^<(\epsilon)=
\frac{i}{2\pi v}
\frac{2^\gamma}{\Gamma(1+\gamma)}
\left(\frac{\pi T}{\Lambda}\right)^\gamma 
e^{-{\pi z\over 2}}
\Bigl|\Gamma[(1+\gamma+iz)/2]\Bigr|^2.
\end{equation}
This yields the following  asymptotic behavior of the Green function at low 
temperatures ($|\epsilon| \gg T$),  
\begin{equation}
G^>(\epsilon)=-\frac{i}{v\Gamma(1+\gamma)}e^{\pi(z-|z|)/2}
\left(\frac{|\epsilon|}{\Lambda}\right)^\gamma\,, 
\end{equation}
and high temperatures ($|\epsilon| \ll T$),
\begin{equation}
G^>(\epsilon)=-\frac{i}{2\pi v}\frac{2^\gamma}{\Gamma(1+\gamma)}
\Gamma^2[(1+\gamma)/2]
\left(\frac{\pi T}{\Lambda}\right)^\gamma\,  \,.
\end{equation} 
Using Eqs.~(\ref{App_nu}), (\ref{App3}), and (\ref{App4}), one obtains
TDOS at equilibrium,
\begin{eqnarray}
\label{App6}
\nu(\epsilon,T)&=&\frac{2^{\gamma-1}}{\pi^2 v\Gamma(1+\gamma)}
\left(\frac{\pi T}{\Lambda}\right)^\gamma \nonumber \\
&\times &
\Biggl|\Gamma[(1+\gamma+iz)/2]\Biggr|^2
\cosh\frac{\pi z}{2}.
\end{eqnarray}
Equation (\ref{App6}) describes the well-known ZBA in TDOS, 
$\nu(\epsilon) \propto |\epsilon|^\gamma$, smeared at the scale
$\epsilon \sim 2\pi T(1+\gamma)$. 
This smearing results from a combined effect of
(i) the thermal broadening of the distribution function and (ii) the
dephasing rate\cite{GGM-2008}  $1/\tau_\phi=2\pi\gamma T$.

It is straightforward to check that the Fermi-Dirac distribution function is 
recovered from the ratio 
\begin{equation}
\frac{G^>(\epsilon)+G^<(\epsilon)}{G^>(\epsilon)-G^<(\epsilon)}=
\tanh\frac{\epsilon}{2T}=1-2n_0(\epsilon),
\end{equation}
in agreement  with  the fluctuation-dissipation theorem.


\begin{thebibliography}{99}
\bibitem{note-relaxation} 
In fact, the word ``relaxation'' should be used with care in the
present situation: as we show, emerging distribution functions
are generically not of equilibrium (Fermi-Dirac) type.
\end{thebibliography}
\end{document}